\documentclass[preprint,usenatbib]{raa}            

\usepackage{graphicx, times}             
\usepackage{lineno}
\usepackage{natbib} 
\usepackage{setspace}

\usepackage[T1]{fontenc}
\usepackage{multicol,multicap,graphicx}
\usepackage{amsmath}	
\usepackage{amssymb}	
\usepackage{subfig}     
\usepackage{breakurl}
\usepackage{float}
\usepackage{flafter}
\usepackage{verbatim}
\usepackage{soul}
\bibpunct{(}{)}{;}{a}{}{,}
\usepackage[final=true,pagebackref=true]{hyperref}
\usepackage{orcidlink}

\begin{document}
    
   \title{Results of 23\,yr of Pulsar Timing of PSR J1453$-$6413
}

   \volnopage{Vol.0 (2023) No.0, 000--000}      
   \setcounter{page}{1}          
   \author {Wei Li~\orcidlink{0009-0009-8247-3576}
    \inst{1,3}
    \and Shi-Jun Dang~\orcidlink{0000-0002-2060-5539}
        \inst{1,2,3}
    \and Jian-Ping Yuan
        \inst {4} 
    \and Lin Li
        \inst{5}
    \and Wei-Hua Wang
        \inst{6}
    \and Lun-Hua Shang
        \inst{1,3}
    \and Na Wang
        \inst {4}
    \and Qing-Ying Li
        \inst{1,3}
    \and Ji-Guang Lu
        \inst {7}
    \and Fei-Fei Kou
        \inst {4}
    \and Shuang-Qiang Wang
        \inst {4}    
     \and Shuo Xiao
        \inst {1,3}
    \and Qi-Jun Zhi
        \inst{1,3}
    \and Yu-Lan Liu
        \inst {2,7,8}
    \and Ru-Shuang Zhao
        \inst{1,3}
    \and Ai-Jun Dong
        \inst{1,3}
    \and Bin Zhang
        \inst{1,3}
    \and Zi-Yi You
        \inst{1,3}    
    \and Yan-Qing Cai
        \inst{1,3}
    \and Ya-Qin Yang
        \inst{1}
    \and Ying-Ying Ren
        \inst{1,3}
    \and Yu-Jia Liu
        \inst{1,3}
    \and Heng Xu
    \inst{1,3}
}

   \institute{{School of Physics and Electronic Science,Guizhou Normal University, Guiyang 550001, China \it dangsj@gznu.edu.cn; qjzhi@gznu.edu.cn; lhshang@gznu.edu.cn}\\ 
        \and
            Guizhou Radio Astronomy Observatory, Guizhou University, Guiyang 550025, China\\
        \and
            Guizhou Provincial Key Laboratory of Radio Astronomy and Data Processing, Guizhou Normal University, Guiyang 550001, China\\
        \and 
            Xinjiang Astronomical Observatory, Chinese Academy of Sciences, Urumqi, Xinjiang 830011, China\\
        \and
            School of Physics Science and Technology, Xinjiang University, Urumqi, Xinjiang 830011, China\\
        \and
            College of Mathematics and Physics, Wenzhou University, Wenzhou 325035, China\\ 
        \and
            National Astronomical Observatories, Chinese Academy of Sciences, Beijing 100101, China\\
        \and 
            CAS Key Laboratory of FAST, National Astronomical Observatories, Chinese Academy of Sciences, Beijing 100101, China\\
        }

\date{Received~~2023 XXXX XXXX; accepted~~XXXX}

\abstract{In this paper, we presented the 23.3 years of pulsar timing results of PSR J1456$-$6413 based on the observation of Parkes 64m radio telescope. We detected two new glitches at MJD 57093(3) and 59060(12) and confirmed its first glitch at MJD 54554(10). The relative sizes ($\Delta\nu/\nu$) of these two new glitches are $0.9 \times 10^{-9}$ and $1.16 \times 10^{-9}$, respectively. Using the “Cholesky” timing analysis method, we have determined its position, proper motion, and two-dimensional transverse velocities from the data segments before and after the second glitch, respectively. Furthermore, we detected exponential recovery behavior after the first glitch, with a recovery time scale of approximately 200 days and a corresponding exponential recovery factor Q of approximately 0.15(2), while no exponential recovery was detected for the other two glitches. More interestingly, we found that the leading component of the integral pulse profile after the second glitch became stronger, while the main component became weaker. Our results will expand the sample of pulsars with magnetosphere fluctuation triggered by the glitch event. 
}
   \authorrunning{Wei Li et al. 2023}            
   \titlerunning{Results of 23 yr of Pulsar Timing of PSR J1453$-$6413}  

   \maketitle

\section{Introduction}\label{sec:introduction}           
Pulsars are rapidly rotating neutron stars with high magnetic fields. Due to the remarkable stability of pulsar rotation, they are considered the ideal clocks. Pulsar timing studies can be carried out with long-term observation records of the arrival time of pulsar pulses to the Earth, and comparison with theoretical models to obtain the arrival time residual, so as to study the physics involved. Pulsar timing can be used for obtaining its rotational and astrometric parameters, detecting the internal states of neutron stars, testing general relativity \citep{2004MNRAS.353.1311H}, detecting gravitational waves \citep{2006Sci...314...97K}, pulsar navigation \citep{2013AdSpR..52.1602D} and pulsar clocks \citep{2023MNRAS.521.2553L}. However, long-term timing observations show that the spin frequency $\nu$ of some pulsars evolves irregularly. Timing noise and glitches are two main manifestations of this irregularity.

Timing noise, which is often presented as non-whiten and un-modeled timing residual, is a significant deviation of the pulsar rotation from the spin-down model indicating erratic fluctuation in the rotation of the pulsar. The phenomenon is classified as low-frequency noise or red noise and usually has time scales of months to years \citep{2010MNRAS.402.1027H}. The nature of timing noise is still unclear, but it has been attributed to either fluctuation in the spin-down or magnetospheric torque \citep{1987ApJ...321..805C, 1987ApJ...321..799C}, or variation of the coupling between the stellar crust and its superfluid core \citep{1990MNRAS.246..364J}. Timing noise has also been proposed that the macroscopic turbulence in the neutron star's core superfluid can lead to instability in the spin-down processes \citep{2014MNRAS.437...21M}. 

A glitch appears as a sudden increase in the pulsar spin frequency. Glitch usually can be divided into large glitch $\sim 10^{-5}$ and small glitch $\sim 10^{-10}$ according to the value of $\Delta\nu/\nu$. Generally, the evolution of $\nu$ of post glitch tends to present exponential recovery, and the spin-down rate $|\dot \nu|$ will increase first and then decay to the level of the pre-glitch. The relaxation time of exponential recovery can vary greatly between pulsars and even after different glitches in the same pulsar \citep{1996Ap&SS.246...73D,2010ApJ...719L.111Y, 2010MNRAS.404..289Y, 2011MNRAS.414.1679E, 2013MNRAS.429..688Y,2020ApJ...896..140D}.

Generally, the glitch is thought to originate inside a neutron star. Two main mechanisms have been proposed to explain this phenomenon. The first mechanism suggests that the crust earthquakes in neutron stars cause an increasing amount of strain to accumulate in the crust, leading to a sudden rearrangement of the moment of inertia \citep{1991ApJ...382..587R, 1998ApJ...492..267R}. But this hypothesis is weak in explaining large glitches. The second mechanism is the sudden transfer of the moment of inertia carried by the superfluid inside the neutron star to its crust, resulting in the observed sudden increase in spin frequency \citep{1975Natur.256...25A, 1976ApJ...203..213R}. This hypothesis successfully explains the post-glitch relaxation of PSR J1453$-$6413 and other pulsars. However, recent observations have shown that glitches in some pulsars are connected with the changes in their pulse profiles, such as PSRs J1119$-$6127, J2036+2740, J0738$-$4042, J0742$-$2822, J0835-4510, B2021+51 and B1822-09 \citep{2011MNRAS.411.1917W, 2018MNRAS.478L..24K, 2021RAA....21...42D, 2023MNRAS.519...74Z, 2013MNRAS.432.3080K,2018Natur.556..219P,2022ApJ...931..103L}. This may imply that the pulsar glitches are related to their magnetosphere activity.

PSR J1453$-$6413 is a normal middle age pulsar with spin period $P \approx$  0.179\,s \citep{2019MNRAS.489.3810P}, and the characteristic age $\tau_{c}\equiv-\frac{\nu}{\left(2\dot\nu\right)} \approx$ 1.037\,Myr, distance $D\approx2.8$\,kpc. Only one glitch of this pulsar at MJD 54552(4) was reported by the previous work, and the relative size of this glitch is $\Delta\nu/\nu\approx2.99(18)\times10^{-10}$ with the relative change of spin-down rate $\Delta\dot\nu/\dot\nu\approx0.55(11)\times10^{-3}$ \citep{2013MNRAS.429..688Y}. 

In this paper, we have studied the timing behavior of PSR J1453$-$6413 using 23.3 years of timing observations of the 64m Radio Telescope (Parkes) in Australia at the central frequency of 1369MHz. We have found two new glitches in this pulsar and obtained the new proper motion and position using the “Cholesky” timing analysis method.  In Section~\ref{sec:obs}, we introduce the observation and data process. In Section~\ref{sec:results}, we report the results of the timing analysis. Discussion and Conclusions follow in Section~
\ref{sec:Discusstion} and Section~\ref{sec:conclusion}.

\section{Observation and Data Analyses}\label{sec:obs}
We analyzed timing observations of PSR J1453$-$6413 from July 1998 to October 2021, which are publicly available to download from the Parkes pulsar data archive\footnote{https://data.csiro.au/dap/public/atnf/pulsarSearch.zul} \citep{2011PASA...28..202H}. The data before MJD 54303 has a center frequency of 1374\, MHz and a bandwidth of 256\, MHz. In addition, the Parkes Analogue Filter Bank system (AFB) was used in the observation during this data segment. The data between MJD 54303 and 58686 has a center frequency of 1369\, MHz and a bandwidth of 256\, MHz, and the multi-beam receiver and the Parkes Digital Filter Bank systems (PDFB1/2/3/4) have been used in the observations. Furthermore, the vast majority of data from MJD 58430 to 59514 was observed with the ultra–wide-bandwidth receiver. The time intervals of observation are usually 2-4 weeks/time, the sub-integration time of each observation is 30 seconds, and the total integration time is 2-15 minutes.

After obtaining the data, we used the software \textsc{PSRCHIVE}\footnote{http://psrchive.sourceforge.net/} \citep{2004PASA...21..302H} and the software packages \textsc{TEMPO2}\footnote{https://bitbucket.org/psrsoft/tempo2/} \citep{2006MNRAS.369..655H} for offline data processing. First, \textsc{PSRCHIVE} was used to remove radio frequency interference (RFI) and incoherent dispersion from data, and time, frequency, and polarization were summed to form an integral pulse profile. Secondly, we select the one with the highest signal-to-noise (S/N) as the standard template and used command \textsc{PAT} in \textsc{PSRCHIVE} to cross-correlational all the average pulse profiles with the standard template to get time-of-arrival (ToA). In order to eliminate the influence of Earth's motion, we transformed these ToAs to the Solar system barycenter (SSB) using the Solar ephemeris DE436 and \textsc{TEMPO2} software packages, which can be regarded as an inertial reference frame. Finally, To obtain the high-precision timing solution, we employ the "Cholesky" timing analysis method to fit the red noise model, which can be achieved by the \textsc{SpectralModel} plugin of \textsc{TEMPO2} software package. The detailed implementation steps of the \textsc{SpectralModel} plugin are seen \citet{2020ApJ...896..140D}. Offset between different observing systems was also included.

\section{Results}\label{sec:results}

\subsection{Position and Proper Motion}

Based on the above methods, we determined the position, proper motion, and transverse velocity of PSR J1453$-$6413. Table~\ref{Table:Position} lists the position of this pulsar, including the positions from the previous publication of the ATNF pulsar catalogue\footnote{http://www.atnf.csiro.au/research/pulsar/psrcat/} \citep{2005AJ....129.1993M} and the parameters obtained by this work. The first and second columns list the Right ascension(RA) and declination(DEC) in J2000 Equatorial Coordinates respectively. The third column lists the reference time for the position. The corresponding data range for each position is listed in the fourth column. The first line shows the results from the literature. The bottom two lines show the results from our timing analyses. Uncertainties in each parameter are in the last quoted digit and are $1\sigma$ (Tables~\ref{Table: Proper motion},~\ref{Table:spinparameters}, and \ref{Table:Glitch parameter} are the same). Compared with the position precision given in the literature, the precision of DEC obtained by the data segment in 57129$-$59035 is higher.
\begin{table}[htb]
\renewcommand{\arraystretch}{1.25}
\caption{The Positions of PSR J1453$-$6413 in J2000 Equatorial Coordinates.}
\label{Table:Position}
    \centering
    \begin{tabular}{ccccc}
    \hline
    RAJ & DECJ & 	POSEPOCH & Data Range & Reference \\
    (h:m:s)	& (d:m:s) & MJD & MJD & 	\\
    \hline 
14:53:32.665(6) & $-$64:13:16.00(5)	 & 55433 &	54220$-$58012 &	\citep{2019MNRAS.489.3810P} \\
14:53:32.663(6) & $-$64:13:15.99(5)	 & 55811 &	54566$-$57057 &	This Work \\
14:53:32.653(1) & $-$64:13:16.070(9) & 58082 &	57129$-$59035 &	This Work  \\
    \hline
    \end{tabular}
\end{table}

\begin{table}[htb]
\renewcommand{\arraystretch}{1.25}
\caption{The Proper motion of PSR J1453$-$6413.}
\label{Table: Proper motion}
    \centering
    \begin{tabular}{cccc}
    \hline
    $\mu_{\alpha}$ & $\mu_{\delta}$ & Data Range & Reference \\
    (mas yr$^{-1}$)	& (mas yr$^{-1}$) & MJD & 	\\
    \hline 
$-$16.4(11) & $-$21.30(76) & 54597$-$57057 & \citep{1990MNRAS.247..322B} \\
$-$14(32)   & 0(41)        & 54597$-$57057 & This Work \\
$-$9(17)    & $-$14(21)    & 57129$-$59035 & This Work  \\
    \hline
    \end{tabular}
\end{table}
Table~\ref{Table: Proper motion} lists the proper motion obtained by our timing analysis and the literature. In our analysis, the parameters of proper motion before and after the second glitch are determined, respectively. The first and second columns list the $\mu_{\alpha}$ and $\mu_{\delta}$, which are projections of proper motion on RA and DEC, respectively. The first line shows the result of previous analyses. The bottom two lines show the results of our timing analyses. 

When analyzing the proper motion of PSR J1453$-$6413, we found that the proper motion of this pulsar changed before and after the second glitch. Due to the impact of this glitch, it was difficult to determine the proper motion in the whole data span. Therefore, we take the epoch of the second glitch as the boundary and get the proper motions of the pre and post-glitch data span, respectively. The values of proper motion in RA and DEC of the pre-glitch are $\mu_{\alpha}$=$-$14(32)\,mas yr$^{-1}$ and $\mu_{\delta}$=0(41)\,mas yr$^{-1}$. The values of proper motion in RAJ and DECJ 
for the post-glitch data span are $\mu_{\alpha}$=$-$9(17)\,mas yr$^{-1}$ and $\mu_{\delta}$=$-$14(21)\,mas yr$^{-1}$. Obviously, our results are well consistent with the literature in the error range.

\subsection{Velocity}

In general, if the proper motion and distance of a pulsar are known, its 2D transverse velocity can be given by $\rm V_{T}$=4.74$\mu_{tot}$$D$\,km s$^{-1}$, where $\rm \mu_{tot}$=$\sqrt{\mu_{\alpha}^{2}+\mu_{\delta}^{2}}$ is the total proper motion in mas yr$^{-1}$, $D$ is the distance of the pulsar in kpc. The uncertainties of the pulsar velocities can be calculated from the standard error transfer formula. 
The distance $D^{Y}=2.8$\,kpc was obtained by measuring the absorption line of HI, and the distance $D=1.432$\,kpc was estimated using the YMW16 Galactic free-electron density model \citep{2017ApJ...835...29Y}. In addition, the uncertainties of the distances $D$ are 40\% \citep{2017ApJ...835...29Y}. We obtained the 2D transverse velocities $V_{T}$ of PSR J1453$-$6413 for the pre and post-glitch data span (see: Table~\ref{Table:Velocity}).

\begin{table}[htb]
\renewcommand{\arraystretch}{1.25}
\caption{The 2D Transverse Velocities of PSR J1453$-$6413.}
\label{Table:Velocity}
    \centering
    \begin{tabular}{cccccc}
    \hline
    $D$ & $V_{T}$ & $D^{Y}$ & $V_{T}^{Y}$ & Data Range & Reference \\
    (kpc)	& (km s$^{-1}$) & (kpc)	& (km s$^{-1}$) & MJD & 	\\
    \hline 
2.2     & 238      & ---        & --- & 54597$-$57057 & \citep{1990MNRAS.247..322B} \\
2.8(11) & 186(431) & 1.4(6) & 95(221)  & 54597$-$57057 & This Work \\
2.8(11) & 221(278) & 1.4(6) & 113(143) & 57129$-$59035 & This Work  \\
    \hline
    \end{tabular}
    \label{tab:my_label}
\end{table}

Table 3 lists the 2D transverse velocity from the previous literature and our results, respectively. The distances are given in the first and third columns, the 2D transverse velocities are given in the second and fourth columns, and data ranges are given in the fifth column. The last column lists the reference. The distance and velocities with superscript “Y” represent those obtained using the YMW16 model. The first line shows the results of the literature. The bottom two lines show the results of our timing analyses. Obviously, these velocity values are also consistent within the range of the error bar.

\subsection{Glitch}

Excepting the first glitch at MJD 54548(10), two new glitches were detected in PSR J1453$-$6413 at MJD 57093(3) and 59060(12). We divided the data span into four segments according to the epoch of these three glitches and then obtained the corresponding rotation parameters ($\nu$ and $\dot\nu$). Uncertainties for the rotation parameters are 1$\sigma$ from \textsc{Tempo2} (see:table~\ref{Table:spinparameters}). 
In addition, we adapt the midpoint of the last observed epoch (T1) before the glitch and the first observed epoch after the glitch as the glitch epoch, and define (T2-T1)/4 as the 1$\sigma$ uncertainty of the glitch epoch\citep{2011MNRAS.414.1679E}. Table 5 shows the $\nu$, $\dot\nu$, and their reference time and the corresponding data range from left to right, respectively.

\begin{table}[htb]
\renewcommand{\arraystretch}{1.25}
\caption{The rotation Parameters of PSR J1453$-$6413}
    \centering
    \begin{tabular}{cllc}
    \hline
    $\nu$ & $\dot\nu$ & Epoch & Data Range \\
    (s$^{-1}$) & 10$^{-14}$\,s$^{-2}$ & (MJD)	& (MJD) 	\\
    \hline 
5.57144894977(4) & $-$8.52013(3)   & 54246 & 50947$-$54549 \\
5.571437431605(6) & $-$8.51892(2)  & 55811 & 54565$-$57057 \\
5.571420718074(1) & $-$8.518586(6) & 58082 & 57129$-$59035 \\
5.57141176670(3)  & $-$8.5203(7)   & 59299 & 59084$-$59514 \\
    \hline
    \end{tabular}
    \label{Table:spinparameters}
\end{table}

\begin{figure}[htb!]
\includegraphics[width=5.5in,height=4.5in,angle=0]{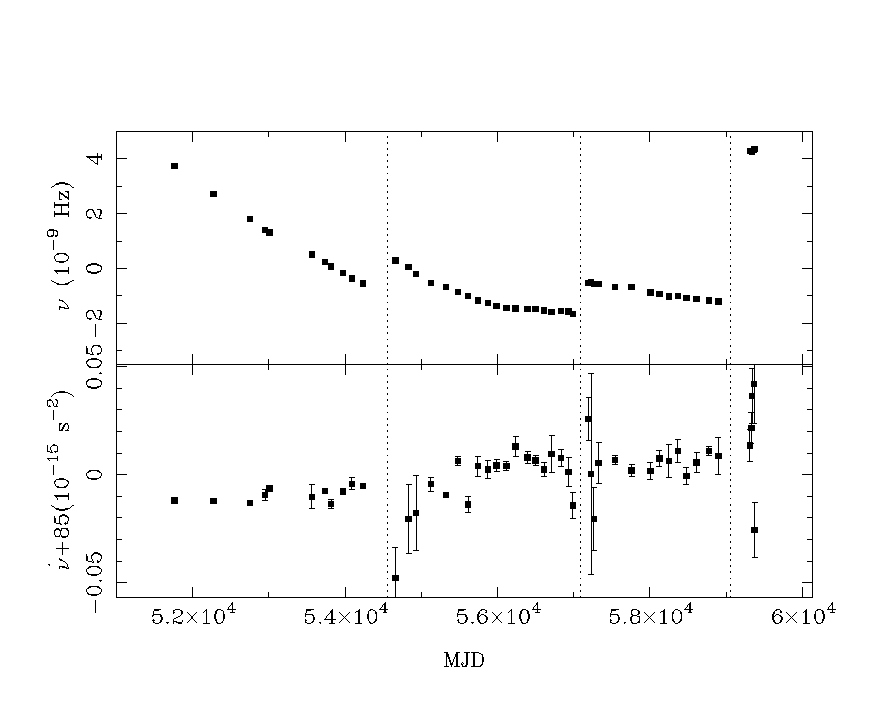}\centering%
\caption{The glitches of PSR J1453$-$6413. The upper panel is $\nu$ with a gradient removed, and the unit is Hz. The bottom panel is $\dot\nu$ with the mean post-glitch values removed, and the unit is s$^{-2}$. Each vertical dashed line presents for the glitch epoch. The horizontal of each panel is time (MJD).}
\label{fig:glitch_evol}
\end{figure}

Figure~\ref{fig:glitch_evol} shows the evolution of $\nu$ and $\dot\nu$ for the three glitches. Obviously, there is a significant recovery process after the first and second glitches, but due to the limited amount of data points, we cannot see its recovery behavior after the third glitch.
In order to quantitatively describe the size of the glitch and the rotation evolution behavior before and after these glitches, the glitch parameters are determined by the \textsc{Tempo2}. The details are as follows:

In the data processing of the glitch of a pulsar, we pay more attention to its rotational phase, which can be expressed by Taylor series expansion,
\begin{equation}
\phi(t)=\phi(t_{0})+\nu(t-t_{0})+\frac{\dot\nu}{2}(t-t_{0})^{2}+\frac{\ddot\nu}{6}(t-t_{0})^{3}+\cdots,
\end{equation}
where $\phi(t_{0})$ is the pulse phase at the barycentric reference time $t_{0}$. Here, $\nu$, $\dot\nu$, and $\ddot\nu$ represent the spin frequency, the first derivative of frequency (the spin-down rate), and the second derivative of frequency respectively. The glitch is identified by a sudden discontinuity in the timing residuals relative to a pre-glitch solution. These pulse phase jumps can usually be described as the increase in the $\nu$ and $\dot\nu$ as follows \citep{2006MNRAS.372.1549E}:
\begin{equation}
\phi_{g}=\Delta\phi+\Delta\nu_{p}(t-t_{g})+\frac{1}{2}\Delta\dot\nu_{p}(t-t_{g})^{2}[1-e^{-(t-t_{0})/\tau_{d}}]\Delta\nu_{d}\tau_{d},
\end{equation}
where $t_{g}$ is the time of glitch, $\Delta\phi$ is the increment of pulse phase in the data before and after the glitch. Glitch is characterized by a permanent increase of spin frequency $\Delta\nu_{p}$, a permanent increase of the first derivative of spin frequency $\Delta\dot\nu_{p}$ and a transient frequency increment $\Delta\nu_{d}$, and the increment decays exponentially on the time scale $\tau_{d}$. Then, increments on the spin frequency and spin-down rate are described by $\Delta\nu_{g}=\Delta\nu_{p}+\Delta\nu_{d}$ and $\Delta\dot\nu_{g}=\Delta\dot\nu_{p}-\frac{\Delta\nu_{d}}{\tau_{d}}$. The exponential recovery factor is defined as $Q\equiv\Delta\nu_{d}/\Delta\nu_{g}$. The glitches parameters of PSR J1453$-$6413 are given in Table~\ref{Table:Glitch parameter}. The errors are derived from the standard error transfer formula.

\begin{table}[]
\renewcommand{\arraystretch}{1.25}
\caption{The Glitch Parameters of PSR J1453$-$6413}
\label{Table:Glitch parameter}
    \centering
    \begin{tabular}{clllllcc}
    \hline
    Gl.No. & Epoch & $\Delta\nu$ & $\Delta\dot\nu$ & $\Delta\nu/\nu$ & $\Delta\dot\nu/\dot\nu$ & $\tau_{d}$ & Q \\
    	& (MJD) & (10$^{-9}$) & (10$^{-15}$) & (10$^{-9}$) & (10$^{-3}$) & (Day) &	\\
    \hline 
1 & 54554(10) & 1.1(2)  & $-$0.03(1)   & 1.9(4)  & 0.3(2) & 200 & 0.15(2) \\
2 & 57093(3)  & 1.16(1) & $-$0.0030(3) & 0.208(3) & 0.035(3) & ---     & ---      \\
3 & 59060(12) & 6.57(4) & $-$0.042(2)    & 1.180(7) & 0.50(2)  & ---     & ---      \\
    \hline
    \end{tabular}
    \label{tab:my_label}
\end{table}

We reconfirmed that the epoch of the first glitch is MJD 54554(10), which is consistent with the previously reported epoch MJD 54552(4). The relative changes in spin frequency and spin-down rate of the first glitch are $1.9(4)\times10^{-9}$ and $0.3(2)\times10^{-3}$ respectively, which are smaller than the previously reported values $0.299(18)\times10^{-9}$ and $0.55(11)\times10^{-3}$, which may be because the previous study did not take the exponential recovery after this glitch into account. The results of our timing analysis show that the epochs for the second and third glitch are MJD 57093(18) and 59060(12) respectively. The relative changes of spin frequency and spin-down rate of the second glitch are $0.208(3)\times10^{-9}$ and $0.035(3)\times10^{-3}$, respectively. And the relative changes in spin frequency and spin-down rate of the third glitch are $1.180(7)\times10^{-9}$ and $0.50(2)\times10^{-3}$ respectively. Obviously, the first glitch is the largest known glitch in PSR J1453$-$6413. However, it should be noted here that we did not consider the exponential recovery behavior in the analysis of the third glitch, so we could not be completely sure of the relative size of this glitch. Our results may need to be revised if exponential recovery behavior exists.

In addition, we also analyzed the post-glitch behavior of three glitches. The result shows that after the first glitch, the pulsar experienced an exponential recovery for about 200 days, with an exponential recovery factor Q $\approx$ 0.15(2). After the second glitch, it shows a linear recovery process. It should be noted that due to the lack of data after the third glitch, we are unable to analyze the post-glitch behavior of the third glitch. With the opening of the subsequent data, we will do further research on the post-glitch behavior of this glitch.

\subsection{Timing noise}
\begin{figure}[htbp]
\centering
\begin{minipage}[t]{0.4\textwidth}
\centering
\includegraphics[width=5cm,angle=0]
{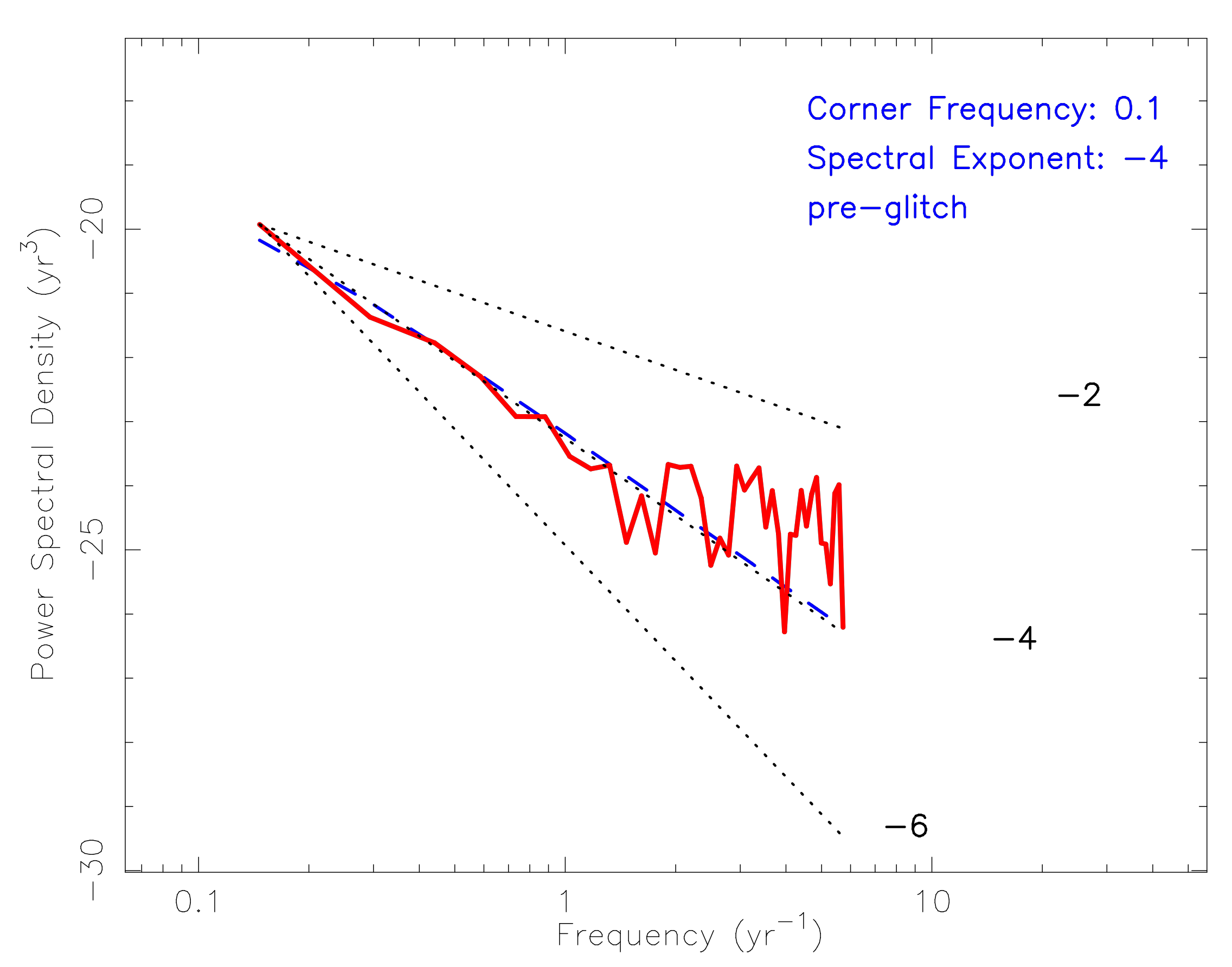}
\end{minipage}
\begin{minipage}[t]{0.4\textwidth}
\centering
\includegraphics[width=5cm,angle=0]{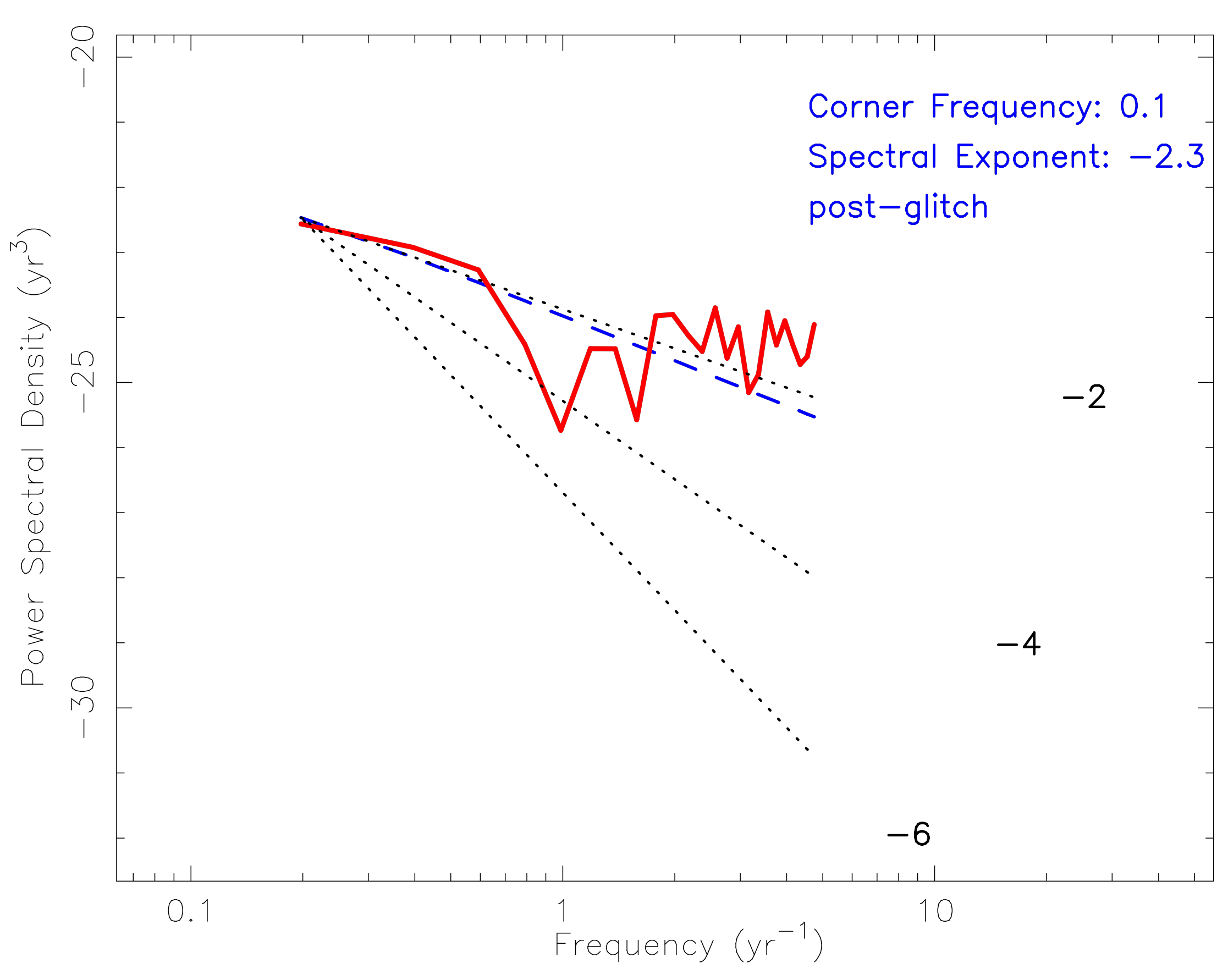}
\end{minipage}
\caption{ Power spectrum of the timing noise in PSR J1453$-$6413 before and after the second glitch. Left: Power spectrum of the pre-glitch timing noise. Right: Power spectrum of the post-glitch timing noise. The red solid lines are the observed spectral noise. The blue dashed lines in each panel represent the red noise model, and the dotted lines stand for the power spectrum with an exponent of $-$2, $-$4, and $-$6 from top to bottom, respectively.}
\label{fig:timing_noise}
\end{figure}

\citet{2011MNRAS.418..561C} developed a method of determining pulsar timing solution by applying the "Cholesky" decomposition on the covariance matrix of timing residual, which is an optimal technique to characterize strong red noise. Since PSR J1453$-$6413 has three glitches within the data span, we divided the data into four sections according to the epoch of glitches. However, because the data span of the first and last sections is insufficient, we only analyzed the timing noise power spectrum of the data span before and after the second glitch (see: Figure~\ref{fig:timing_noise} ).
The power spectrum can be modeled by a power law $P(f)=A/[1+(f/f_{c})^{2}]^{\alpha/2}$, where $A$, $f_{c}$, and $\alpha$ represent amplitude, corner frequency, and spectral index, respectively \citep{2011MNRAS.418..561C}.
The dotted lines in Figure~\ref{fig:timing_noise} indicate the spectral exponent of $-$2, $-$4, and $-$6, which suggest that the noise is dominated by a random walk in the phase, $\nu$, and $\dot\nu$, respectively \citep{1972ApJ...175..217B}. 
For the pre-glitch data, a mode with $\alpha=-4$ and $f_{c}=0.1$\,yr$^{-1}$ could nicely describe the power spectrum of the timing noise. And for the post-glitch data, these values are $\alpha=-2.3$ and $f_{c}=0.1$\,yr$^{-1}$ for a reasonable fit. These results show that the red noise before the second glitch is dominated by random walk in $\nu$, while the red noise after this glitch the spectral index of which is close to $-$2 may be dominated by random walk in the phase.
Our results indicate that the red noise of the pre-glitch data span may be dominated by the random walk in $\nu$, while the red noise of the post-glitch data span may be dominated by random walk in the phase.


\subsection{Correlation between Pulse Profile Changes and Glitches}

\begin{figure}[htb!]
\includegraphics[width=5.7in,height=4in,angle=0]{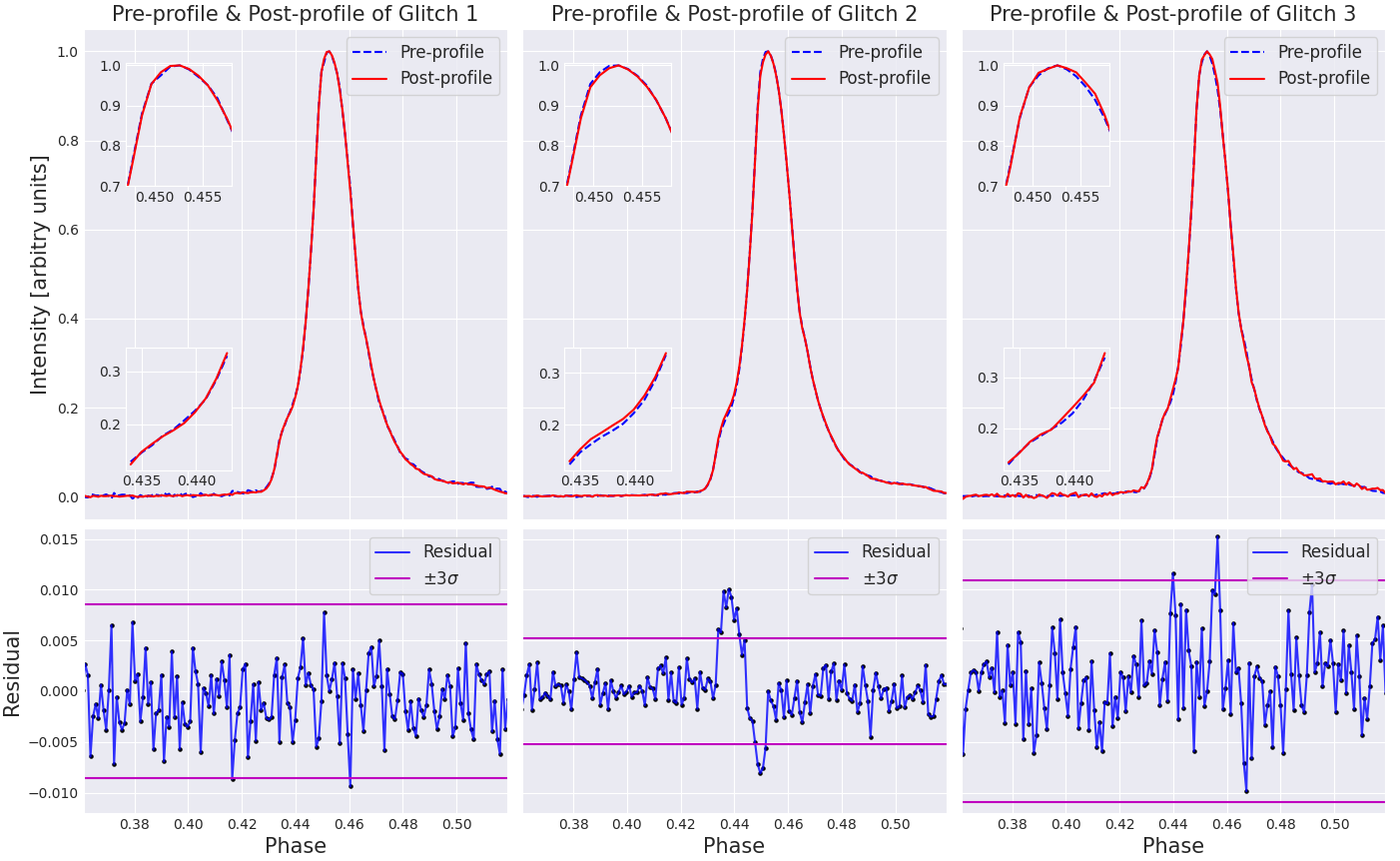}\centering%
\caption{The pulse profile change around three glitches. In the top three panels, the blue dotted lines represent the normalized pulse profile before the glitch, the red solid lines represent the normalized pulse profile after the glitch. The inserts in each panel are the extent of pulse profiles near the 20\% height of the pulse peak and the extent near the pulse peak. The bottom three panels show the residual of the pre- and post-glitch profiles. The purple solid lines in the three bottom panels represent the 3 times the maximum root mean square value of the profile residuals before and after the glitch.}
\label{fig:Profilechange}
\end{figure}


In order to see if there is a correlation between glitches and profiles in PSR J1453$-$6413, we divided the data into four sections according to the epoch of three glitches and obtained the normalized mean pulse profiles before and after each glitch. It should be noted that due to the evolution of pulse profiles at different frequencies, the normalized mean pulse profile before the first glitch does not contain the data with a center frequency of 1374\,MHz.
Figure~\ref{fig:Profilechange} shows the pulse profile change around three glitches. In the top three panels, the blue dotted lines represent the normalized pulse profiles before the glitch, the red solid lines represent the normalized pulse profiles after the glitch. The inserts in each panel are the extent of pulse profiles near the 20\% height of the pulse peak and the extent near the pulse peak. The bottom three panels show the residual of the pre- and post-glitch profiles. The purple solid lines in the three bottom panels represent the 3 times the root mean square value of the profile residuals before and after the glitch. Obviously, there was no significant change in the integral pulse profiles before and after the first and third glitches, while there was a significant change in the pulse profiles before and after the second glitch, with the residual of the leading part and the residual at the peak significantly exceeding three times of its root mean square value. 
The leading component became stronger and the main component became weaker after the second glitch. 
Alternatively, the pulse profile widened near the height of 20\% of the pulse peak, but narrowed in front of the peak after the second glitch. 
However, due to the lack of data after the third glitch, we could not completely determine the correlation between the third glitch and the pulse profile change. With the release of subsequent data, we will further study the correlations between the glitch and the pulse profile change of PSR J1453$-$6413.

\section{Discussion}\label{sec:Discusstion}

\subsection{Proper Motion}
Pulsars are fast-moving objects whose space velocities are an order of magnitude larger than those of their progenitors \citep{1970ApJ...160..979G, 2005MNRAS.360..974H}. A natural explanation for such high velocities is that this is the result of a moment kick at birth \citep{1998ApJ...505..844L, 2001ApJ...549.1111L, 2006ApJ...644..445N}. Velocities of pulsars are determined by measuring their proper motion and distance D. For high-intensity radio pulsars, Very Long Baseline Interferometry (VLBI) can be used to obtain their location and proper motion \citep{2002ApJ...571..906B, 2003AJ....126.3090B, 2016ApJ...828....8D}. In recent years, VLBI has also obtained the proper motion of some weak pulsars, for example, Yan et al. (\citeyear{2013MNRAS.433..162Y}) used Very Long Base Array (VLBA) and European VLBI Network (EVN) to measure astrometric parameters of PSR B1257+12 in a three-planet pulsar system, and Du et al. (\citeyear{2014ApJ...782L..38D}) used EVN to obtain parallax and proper motion of weak millisecond pulsar PSR J0218+4232 at 1.6GHz. In addition, astrometric parameters can be determined by timing observations of pulsars spanning several years \citep{1974ApJ...189L.119M, 2005MNRAS.360..974H, 2016MNRAS.460.4011L, 2020ApJ...896..140D}. In this paper, the position and proper motion of PSR J1453$-$6413 before and after the second glitch were obtained by using pulsar timing analysis. Although the accuracy of our results is not as high as that of the references due to the influence of red noise and the proper motion parameters being different before and after the second glitch, they are consistent within the error range.

\begin{figure}[htb!]
\includegraphics[width=4.5in,height=3.5in,angle=0]{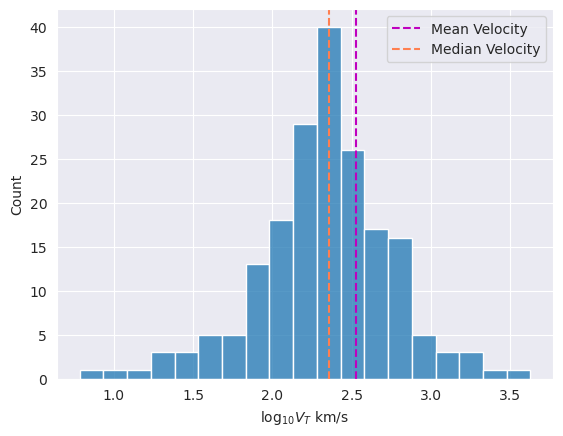}\centering%
\caption{Distribution of the 2D transverse velocity. The longitudinal axis is the count. The horizontal axis is $log_{10} V_{T}$, and the unit is km/s. The two vertical lines with different colors represent the median velocity and mean velocity, respectively.}
\label{fig:glitch}
\end{figure}

The 2D transverse velocities of PSR J1453$-$6413 before and after the second glitch are 186(431)\,km s$^{-1}$ and 221(278)\,km s$^{-1}$, respectively. In order to compare the 2D transverse velocities of other pulsars, we used the data from the ATNF pulsar catalog to count the 2D transverse velocities of normal single pulsars. In this sample including 191 pulsars, the rotation period $P$ of these pulsars is larger than 0.01s, and pulsars in binary systems were excluded. First, we used the distances which are best estimates using the YMW16 DM-based distance as default to count the velocities. Our statistical results show that 2D transverse velocities of normal single pulsars in our sample are unimodal, the mean 2D transverse velocity is 335(38)\,km $^{-1}$, and the median 2D transverse velocity is 229\,km s$^{-1}$. In addition, the highest 2D speed in our sample is 4250(3607)\,km s$^{-1}$ for PSR J1829$-$1751, and the lowest 2D speed is 6(5)\,km s$^{-1}$ for PSR J1752$-$2806. Compare with our statistical result, the 2D transverse velocities of PSR J1453$-$6413 before and after the second glitch are consistent with the mean value within the margins of errors.

Moreover, Hobbes et al. (\citeyear{2005MNRAS.360..974H}) counted the proper motion of 233 pulsars. Their results show that the mean 2D transverse velocities of normal single pulsars are 211(18)\,km s$^{-1}$ with the CL02 model \citep{2002astro.ph..7156C} and 269(25)\,km s$^{-1}$ with the TC93 model \citep{1993ApJ...411..674T}. Inspired by them, we carried out simple statistics on the 2D transverse velocities of normal single pulsars, but the distances of these pulsars are all based on the YMW16 model. Our result shows that the 2D transverse velocities of normal single pulsars based on the YMW16 model are also unimodal, the mean value is 395(44)\,km s$^{-1}$, the median value is 213\,km s$^{-1}$, the highest speed is 5035(596)\,km s$^{-1}$ for PSR J2346$-$0609, and the lowest speed is 7(9)\,km s$^{-1}$. 
The difference in average two-dimensional transverse velocities between these two types may mainly come from the differences in the electron density models of the Galactic.

\begin{figure}[htb!]
\includegraphics[width=4.5in,height=3.5in,angle=0]{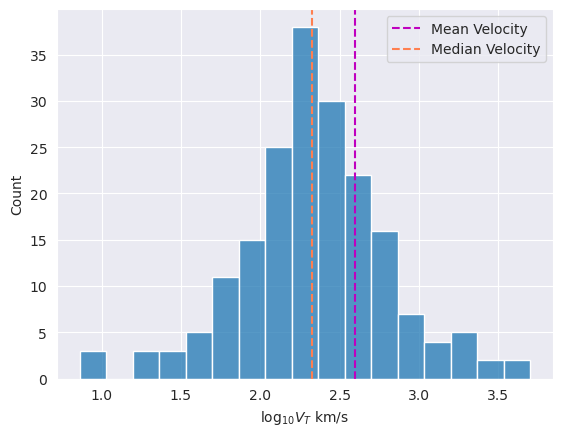}\centering%
\caption{Distribution of the 2D transverse velocity based on the YMW16 model. The longitudinal axis is the count. The horizontal axis is $log_{10} V_{T}$, and the unit is km/s. The two vertical lines with different colors represent the median velocity and mean velocity, respectively.}
\label{fig:glitch}
\end{figure}

\subsection{Glitch}
We detected three glitches in PSR J1453$-$6413 using Parkes 64m radio telescope. And the last two glitches are newly discovered in this study. These three glitches are all very small, which respectively have fractional sizes of $1.9(4)\times10^{-9}$, $0.208(3)\times10^{-9}$ and $1.180(7)\times10^{-9}$. In the framework of the vortex model, the layer between the rapidly rotating superfluid and the shell leads to the Magnus force of crustal stress. When this force exceeds a critical threshold, the vortex will suddenly stop fixing. As a result, angular momentum is transferred to the crust causing a glitch.

If we assume that this pulsar star has only had three glitches, then we can obtain its glitch activity parameter, which is defined as
\begin{equation}
A_{g}=\frac{1}{T}\sum\frac{\Delta\nu_{g}}{\nu},
\end{equation}
where $T$ is the total data span \citep{1990Natur.343..349M, 2000MNRAS.315..534L}. For PSR J1453$-$6413, the corresponding value of $A_{g}\approx1.41\times10^{-10}$\,yr$^{-1}$. This is consistent with the assumption that the low glitch activity $A_{g}$ of pulsars with small spin-down rate is caused by small and rare glitches \citep{2011MNRAS.414.1679E, 2010MNRAS.402.1027H, 2000MNRAS.315..534L, 2010MNRAS.404..289Y}. \citet{2011MNRAS.414.1679E} reported a correlation between the mean glitch rate and spin-down rate with $\left\langle N_{g} \right\rangle\propto|\dot\nu|^{0.47(4)}$. This result suggests that pulsars with low spin-down rates tend to exhibit small glitches. It should be noted that this correlation is based on a large sample, those pulsars that did not be detected with glitches were included and some newly detected glitches were not included. For PSR J1453$-$6413 with $|\dot\nu|\approx85.1812\times10^{-15}$\,s$^{-2}$, the predict $\left\langle N_{g} \right\rangle\ $ is in the range of 0.027(5) to 0.04(1)\,yr$^{-1}$. 
The observed mean glitch rate is large with the value of $\approx0.13$\,yr$^{-1}$, which is about 3.3 to 4.8 times that of the predicted value. This suggests that, in terms of glitch, PSR J1453$-$6413 is more active than those with similar spin-down rates.

We investigated radio Pulsar with characteristic ages between $1.0\times10^{6}$ and $1.5\times10^{6}$\,yr, and found that only 11 pulsars were detected to have undergone glitches, with a total glitch number of 30\footnote{https://www.atnf.csiro.au/people/pulsar/psrcat/glitchTbl.html} $^{,}$ \footnote{https://www.jb.man.ac.uk/pulsar/glitches/gTable.html}. These pulsars including PSR J1453$-$6413, and six pulsars (PSRs J0528+2200, J1453$-$6413, J1705$-$1906, J2225+6535, J1812$-$1718) have each at least three glitches observed \citep{2011MNRAS.414.1679E, 2010MNRAS.404..289Y, 2006A&A...457..611J, 2013MNRAS.429..688Y, 2022MNRAS.510.4049B, 2018Ap&SS.363...96L, 1982ApJ...255L..63B}. For those radio pulsars with similar period $P$ (from 0.17\,s to 0.19\,s) and similar $\dot P$ (from $1\times10^{-14}$ to $1\times10^{-13}$), no glitch was observed expected PSR J1453$-$6413. Almost all glitches of these pulsars with similar characteristic ages and spin parameters have relative size of $\approx10^{-9}$, except for the glitches in PSRs J0611+1436 and J2225+6535, which have relative glitch sizes of $\approx10^{-6}$ \citep{2022MNRAS.510.4049B, 1982ApJ...255L..63B}. Therefore, we could not rule out the possibility of large glitches happening to PSR J1453$-$6413.

There are two popular theories to explain the glitches of pulsars. The first theory holds that the glitch is caused by starquakes, and the second theory holds that the glitch is caused by the interaction between the superfluid inside the pulsar and the solid crust. The former can only explain the glitch of a few pulsars \citep{1991ApJ...382..587R,1998ApJ...492..267R}, while the latter is the mainstream theory at present, which can be used to explain the glitch of most pulsars \citep{1975Natur.256...25A,2014PhRvC..90a5803P}. The coupling parameters between the superfluid inside the pulsar and the outer crust are defined as $G=\tau_{c}A_{g}=\frac{\overline{\nu}}{|\dot\nu|}\frac{1}{T}\sum\frac{\Delta\nu_{g}}{\nu}$. Here $\frac{I_{c}}{I}\geq G$, $I_{c}$ is the moment of inertia of the crust superfluid and $I$ is the total moment of inertia. For Vela-like mature pulsars which glitch frequently and experience large glitches, if we consider the exit of the crustal entrainment, the lower limit of $\frac{I_{c}}{I}$ is $1.6\% \sim 7\%$ \citep{2014PhRvC..90a5803P}. More than 7\% will result in the superfluid in the crust cannot meet the angular momentum required by the glitch, and other possibilities such as core superfluid involvements should be considered \citep{2015SciA....1E0578H}. For PSR J1453$-$6413, the corresponding $\frac{I_{c}}{I}$ is about 0.029\%, which lies within the theoretical expectations of superfluid theories \citep{2014PhRvC..90a5803P}. Its small value of $\frac{I_{c}}{I}$ could result from the non-detection of large glitches in PSR J1453$-$6413 during our current observational time span. In other words, if another glitch of magnitude $10^{-8}$ should be found in the future, the fractional moment of inertia $\frac{I_{c}}{I}$ will increase by one order of magnitude and return to normal.


\subsection{Correlation between Pulse Profile Changes and Glitches}
Increasing evidence shows that there seem to be some complex correlations between pulsars’ spin and emission. However, we could not fully understand these correlations at present. Nevertheless, the view that the measurable changes in spin-down rate, flux, and pulse shape of pulsars are driven by a shift in the magnetic inclination angle $\alpha$ as a consequence of a glitch, has gradually become a consensus \citep{1997ApJ...478L..91L, 2015MNRAS.449..933A, 2016ApJ...825...18N, 2021ApJ...912...58L}. In general, the glitch is thought to originate inside the neutron star, but the emission is thought to originate in the magnetosphere of the neutron star. The phenomenon of emission changed by glitches has provided an opportunity for us to study the interaction between spin and emission. The change of $\Delta\dot\nu$ reflects the change of external braking torque of the pulsar. The permanent relative change of $\Delta\dot\nu$ can be caused by change in inclination angle \citep{1992ApJ...390L..21L, 1997ApJ...478L..91L}. According the MHD simulation of Spitkovsky (\citeyear{2006ApJ...648L..51S}), the relationship between the magnetic angle $\alpha$ and change of spin-down rate $\Delta\dot\nu/\dot\nu$ can be expressed as follows: $\frac{sin2\alpha\Delta\alpha}{(1+sin^{2}\alpha)^{2}}$. Glitch may lead to the change of the pulsar magnetic field structure and the inclination angle \citep{2016ApJ...825...18N}. Corresponding to the increase of 0.0035\% in $\Delta\dot\nu/\dot\nu$ of PSR J1453$-$6413, the expected change in inclination angle is $\Delta\alpha\approx0.003\,^{\circ}$, if we assume the inclination angle for this pulsar is 45\,$^{\circ}$. The change in inclination angle will cause the change of effective emission geometry, which will result in the change of pulse profile we observed. Here we can see that change in the inclination angle of this pulsar is very small, which is consistent with the small change in the pulse profile we observed.

\section{Summary}\label{sec:conclusion}
We have presented a timing analysis of PSR J1453$-$6413 based on 23.3 yr of observations using the Parkes 64-m radio telescope. The main conclusions are as follows: 
\begin{enumerate}
\item Using the Colesky method, we obtained the improved values of position, proper motion, velocity, and spin parameters of this pulsar. 
\item We confirmed the first glitch of this pulsar in MJD 54554(10) and detected an exponential recovery of this glitch, with a time scale of about 200 days and the corresponding Q factor is about 0.15. Besides, we also detected two new glitches that occurred in MJD 57093(18) and 59060(12), respectively. The relative sizes of these three glitches are $1.9(4)\times10^{-9}$, $0.208(3)\times10^{-9}$ and $1.180(7)\times10^{-9}$, respectively.  
\item We obtained the power spectrum of the timing noise of this pulsar, and found that the red noise before the second glitch may be dominated by the random walk in $\nu$, while the red noise after the second glitch may be dominated by the random walk in the pulse phase.
\item  We detected the glitch correlated profile changing of the second glitch. The leading component of the integral pulse profile after the second glitch became stronger, while the main component became weaker. 
\end{enumerate} 
We look forward to collecting more observations in the future to further investigate the correlation between the glitches and pulse profile changes of PSR J1453$-$6413.

\section{Acknowledgements}
This work was supported by the National SKA Program of China (No.2022SKA0130100, 2020SKA0120100, 2022SKA0130104), Guizhou Province Science and Technology Foundation (No. ZK[2022]304), the Major Science and Technology Program of Xinjiang Uygur Autonomous Region (Nos. 2022A03013-2, 2022A03013-4), the Scientific Research Project of the Guizhou Provincial Education (Nos. KY[2022]132, KY[2022]123, KY[2022]137), the National Natural Science Foundation of China (Nos.11873080, U1731238, 11565010, 12103013, U1838109, U1831120, 12273008, 12103013), the Joint Research Fund in Astronomy under cooperative agreement between the National Natural Science Foundation of China and Chinese Academy of Sciences (No. U1931101), the Foundation of Guizhou Provincial Education Department (No. KY(2020)003, KY(2021)303), the Guizhou Province Science and Technology Support Program (No. [2023] General 333), the 2021 project Xinjiang Uygur autonomous region of China for Tianshan elites, the Key Laboratory of Xinjiang Uygur Autonomous Region No. 2020D04049, and the CAS Jianzhihua project. 
The Parkes radio telescope is part of the Australia Telescope National Facility which is funded by the Commonwealth of Australia for operation as a National Facility managed by CSIRO. This paper includes archived data obtained through the CSIRO Data Access Portal.

\bibliographystyle{raa}

\bibliography{mybibfile.bib}

\label{lastpage}

\end{document}